\documentclass[10pt,twocolumn,english,prd,superscriptaddress,nofootinbib,preprintnumbers,showpacs,floatfix]{revtex4-2}

\usepackage[utf8]{inputenc}
\usepackage{amsmath}
\usepackage{amsfonts}
\usepackage{amssymb}
\usepackage{graphicx} 
\usepackage[caption=false]{subfig}
\graphicspath{ {figures/} }
\usepackage[usenames,dvipsnames]{xcolor}
\usepackage{hyperref}   
\hypersetup{
    colorlinks=true, 
    citecolor=MidnightBlue,
    linkcolor=MidnightBlue,
    urlcolor=Cyan}
\usepackage{tikz}
\usepackage{verbatim} 
\usepackage{soul} 

\definecolor{purple}{rgb}{1,0,1}
\definecolor{lime}{HTML}{A6CE39} % needs xcolor

\begin{document}

\title{The Emergence of Measured Geometry in Self-Gravitating Systems}

\author{Maria I. R. Lourenço}
\email{fc56407@alunos.fc.ul.pt}
\affiliation{Instituto de Astrof\'{i}sica e Ci\^{e}ncias do Espa\c{c}o, Faculdade de Ci\^{e}ncias da Universidade de Lisboa, Edifício C8, Campo Grande, P-1749-016 Lisbon, Portugal}

\author{Julian Barbour} 
\email{julian.barbour@physics.ox.ac.uk}
\affiliation{College Farm, The Town, South Newington, Banbury, OX15 4JG, UK}

\author{Francisco S. N. Lobo} 
\email{fslobo@ciencias.ulisboa.pt}
\affiliation{Instituto de Astrof\'{i}sica e Ci\^{e}ncias do Espa\c{c}o, Faculdade de Ci\^{e}ncias da Universidade de Lisboa, Edifício C8, Campo Grande, P-1749-016 Lisbon, Portugal}
\affiliation{Departamento de F\'{i}sica, Faculdade de Ci\^{e}ncias da Universidade de Lisboa, Edif\'{i}cio C8, Campo Grande, P-1749-016 Lisbon, Portugal}

\date{\LaTeX-ed \today}

\begin{abstract}

This work investigates the geometrical properties of self-gravitating \(N\)-body systems from the perspective established by Henri Poincaré and Albert Einstein concerning the operational nature of measured geometry. Utilizing recent numerical analyses of central configurations---special equilibrium solutions to the Newtonian \(N\)-body problem---we uncover systematic spatial variations in nearest-neighbor particle separations correlated with the radial distance from the system's center of mass. We argue that these variations reflect a context-dependent, emergent effective geometry shaped by gravitational interactions, in accordance with Poincaré's assertion that measured geometry depends on the forces influencing measuring devices, and Einstein's view that rods and clocks define physical geometry through their local dynamics. By revisiting these foundational insights within a modern computational framework, we provide evidence that geometry in self-gravitating Newtonian systems is not a fixed background, but an emergent construct arising from internal physical interactions.
\end{abstract}

\maketitle

\def\HMS{{\scriptscriptstyle{\rm HMS}}}

%\tableofcontents

\section{Introduction}

The understanding of space and geometry as elements of physical reality has undergone a fundamental transformation from classical mechanics to modern physics. Within the Newtonian framework, space is treated as an absolute and immutable Euclidean background in which lengths and time intervals remain invariant under coordinate transformations \cite{Newton1687}. This conception was critically reexamined in the early twentieth century through the philosophical and physical analyses of Henri Poincaré and Albert Einstein. Essentially, concerning the operational nature of measured geometry, they argued that geometry should cease to be a purely abstract mathematical construct and become an operational, phenomenological description shaped by the interaction between matter and measurement.

By measured geometry we mean the effective spatial geometry inferred from the outcomes of physical measurements performed with material devices (rods, clocks, or, in our discrete setting, nearest-neighbour separations between particles). It contrasts with background geometry -- the mathematical space (e.g., Euclidean space) assumed in the formulation of a theory. Measured geometry is operational: its properties depend on the forces acting on the measuring apparatus. In a self-gravitating system, the gravitational field influences inter-particle distances, leading to a position-dependent effective metric, i.e., a non-trivial measured geometry even when the background is flat.

Within this framework, central configurations of the Newtonian $N$-body problem offer a natural setting for examining the physical realization of geometry. Recent numerical studies of central configurations exhibit a clear and robust trend: the typical nearest-neighbor separation tends to increase with distance from the center of mass. This radially dependent pattern reflects a pronounced spatial inhomogeneity in particle separations. If nearest-neighbor separations are interpreted as the lengths of idealized local ``measuring rods'', these results imply that the effective geometry sampled within the system varies with position. In this sense, the observed inhomogeneities provide concrete evidence that geometry in self-gravitating systems is an emergent, context-dependent construct shaped by local interactions.

Contemporary theoretical frameworks resonate with this perspective. In particular, Newton--Cartan theory~\cite{malament2012} and emergent gravity proposals, notably those advanced by Verlinde~\cite{verlinde2011,verlinde2017}. Motivated by these perspectives, the present work analyzes numerical realizations of central configurations to demonstrate how inter-particle separations naturally encode an emergent and spatially inhomogeneous geometry. By situating these results within the conceptual frameworks articulated by Poincaré and Einstein, we clarify how measured geometry in classical self-gravitating systems arises operationally from discrete matter distributions and their mutual interactions.

This work is organized in the following manner: in Sec.~\ref{secII}, we introduce the concept of variety as a scale-invariant measure on shape space. Section~\ref{secIII} analyzes central configurations of the Newtonian $N$-body problem and shows how spatial inhomogeneities in inter-particle separations give rise to an effective, position-dependent measured geometry. Section~\ref{secIV} situates these results within the relational and operational perspectives associated with Henri Poincar\'e and Albert Einstein. Section~\ref{secV} develops a unified framework for emergent measured geometry in gravitational systems, and Sec.~\ref{secVI} concludes with a summary and discussion of implications and future directions.

\section{Variety}\label{secII}

Consider a system of $N$ point particles, each with mass $m_i$, such that $M = \sum_i m_i$, and positions $\mathbf{r}_i$ in Cartesian coordinates. 
We aim to define a scale-invariant function that measures the degree of uniformity or clustering of the distribution. We will call it the \emph{variety} \cite{paper1}.

A natural approach, given that our ontology is based on particles and their mutual distances, is to construct a dimensionless quantity from the ratio of two length measures. We select the \emph{root-mean-square length}, 
\begin{equation}
	\ell_{rms} = \frac{1}{M} \sqrt{\sum_{i<j} m_i m_j r_{ij}^2},
\end{equation}
and the \emph{mean-harmonic length}, 
\begin{equation}
	\ell_{mhl}^{-1} = \frac{1}{M^2} \sum_{i<j} \frac{m_i m_j}{r_{ij}},
\end{equation}
where $r_{ij} = \|\mathbf{r}_i - \mathbf{r}_j\|$.

The variety, $V$, is then defined as the ratio of these two scales:
\begin{equation}
	V := \frac{\ell_{rms}}{\ell_{mhl}}.
\end{equation}

Both $\ell_{rms}$ and $\ell_{mhl}$ carry additional physical significance. Specifically, $\ell_{rms}^2$ is proportional to the center-of-mass moment of inertia $I_{cm}$, representing the overall size of the system, while $\ell_{mhl}$ is proportional to the negative inverse of the Newtonian potential $V_{New}$ (with $G=1$). Consequently, $V$ can also be expressed as
\begin{equation}
	V = - \frac{1}{M^{5/2}} \sqrt{I_{cm}} V_{New}.
\end{equation}

In $N$-body theory, $V$ is sometimes referred to as the \emph{shape potential} or \emph{normalized Newton potential}. It is highly sensitive to clustering: when a few particles come very close together, $\ell_{rms}$ remains largely unchanged, but $\ell_{mhl}$ decreases significantly, producing a corresponding increase in $V$. Remarkably, this purely mathematical construct not only quantifies structural variety but also governs gravitational interactions.

Furthermore, $V$ encodes an intrinsic scale for the particle ensemble. While $\ell_{rms}$ reflects the average of the larger separations and $\ell_{mhl}$ the average of the smaller ones, their ratio provides a natural length scale internal to the system. No external reference, such as a fixed ruler, is required.

It could be argued that combining $\sqrt{I_{cm}}$ and $V_{New}$ into a single quantity is somewhat artificial, since they originate from distinct physical considerations. However, from the perspective of a relational, dimensionless characterization, it is precisely $V$---not its individual components---that represents the physically meaningful measure. The separation into $\sqrt{I_{cm}}$ and $V_{New}$ is merely conventional.

\section{Central Configurations and Spatial Inhomogeneity in Particle Separations}\label{secIII}

Central configurations (CCs) occupy a distinguished position in the theory of the Newtonian $N$-body problem. They are defined as configurations of point particles for which the total gravitational force acting on each particle is proportional to its position vector with respect to the center of mass of the system. Explicitly, for particles of masses $m_i$ located at positions $\mathbf{r}_i$, a configuration is central if there exists a constant $\lambda > 0$ such that
\begin{equation}
	\sum_{j \neq i} G \frac{m_j(\mathbf{r}_j - \mathbf{r}_i)}{\|\mathbf{r}_j - \mathbf{r}_i\|^3}
	= -\lambda(\mathbf{r}_i - \mathbf{r}_{cm}),
\end{equation}
for all $i = 1, \dots, N$, where $\mathbf{r}_{cm}$ denotes the center of mass. This condition ensures that the configuration evolves under the equations of motion either by uniform scaling (homothetic solutions) or by rigid rotation (relative equilibria) \cite{moulton1910, saari_cc, meyer2009}.

Historically, central configurations have been studied primarily for their dynamical significance in celestial mechanics, including their role in total-collision solutions, Lagrange equilibria and the qualitative organization of phase space \cite{saari_cc}. However, beyond their dynamical utility, central configurations provide a uniquely controlled setting in which the intrinsic spatial organization of self-gravitating systems can be examined independently of time evolution. Since CCs are equilibria in shape space---the space of configurations modulo translations, rotations and dilatations---they allow one to isolate purely geometric features induced by gravity from transient dynamical effects.

Our numerical investigations of three-dimensional central configurations with equal masses reveal a systematic and robust spatial inhomogeneity in particle separations. In particular, the distribution of nearest-neighbor (NN) distances exhibits a clear dependence on the radial position of particles relative to the center of mass. Particles located closer to the center are surrounded by significantly shorter nearest-neighbor separations, whereas particles at larger radii exhibit larger typical separations. This radial gradient persists across a wide range of particle numbers and numerical realizations, indicating that it is not a finite-size artifact but an intrinsic property of gravitational equilibrium configurations.
All numerical results presented in this section correspond to such stationary shape-space equilibria, so time evolution does not affect the reported nearest-neighbour statistics.

\begin{figure}[ht!]
	\centering
	\includegraphics[width=\columnwidth]{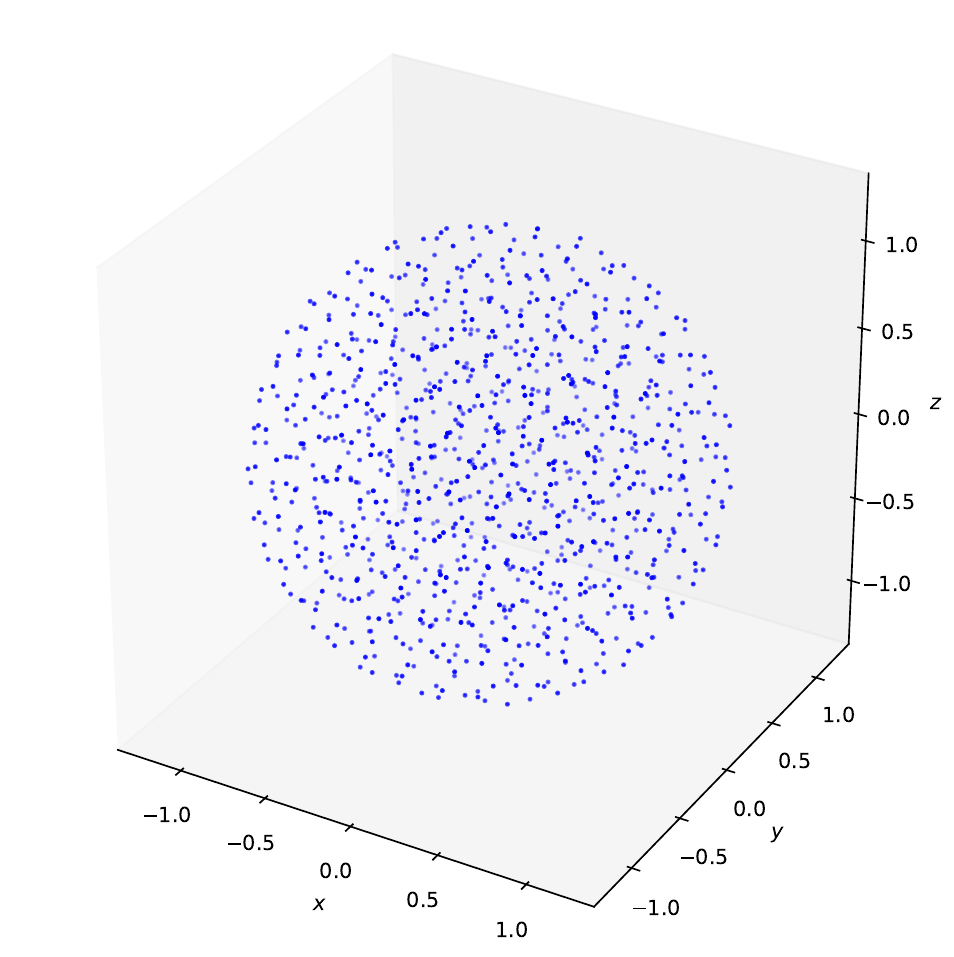}
	\caption{Three-dimensional CC of 1000 equal-mass particles with variety very close to the absolute minimum of $V$
		 ($V_{\text{min}} = 0.457916$).}
	\label{fig1}
\end{figure}

\begin{figure}[ht!]
	\centering
	\includegraphics[width=\columnwidth]{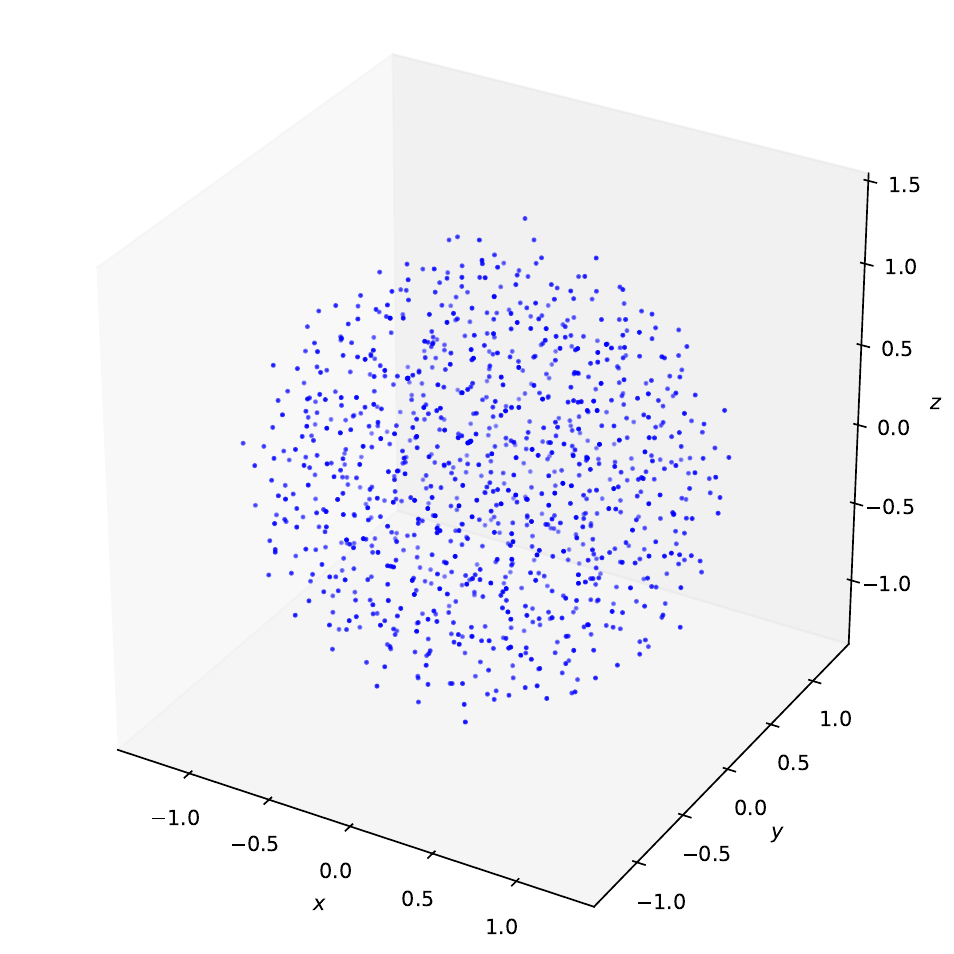}
	\caption{Three-dimensional CC of 1000 equal-mass particles with variety about $1.5\%$ above the absolute minimum of $V$ (Variety $V = 0.465304$, with $1.5\%$ above $V_{\text{min}}$).}
	\label{fig2}
\end{figure}

\begin{figure}[ht!]
	\centering
	\includegraphics[width=\columnwidth]{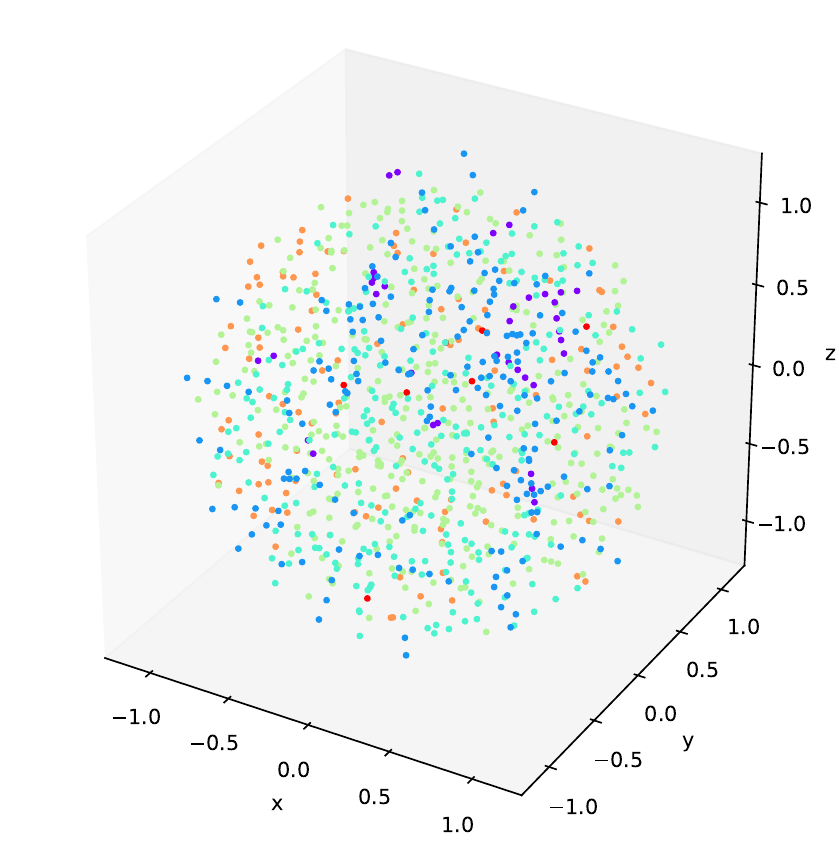}
	\caption{The previous 1000-particle CC roughly $1.5\%$ above the absolute minimum of $V$. We applied a rainbow color scale,
		where particles shift from red to purple as their distance to the nearest-neighbor decreases.}
	\label{fig3}
\end{figure}

\begin{figure}[ht!]
	\centering
	\includegraphics[width=\columnwidth]{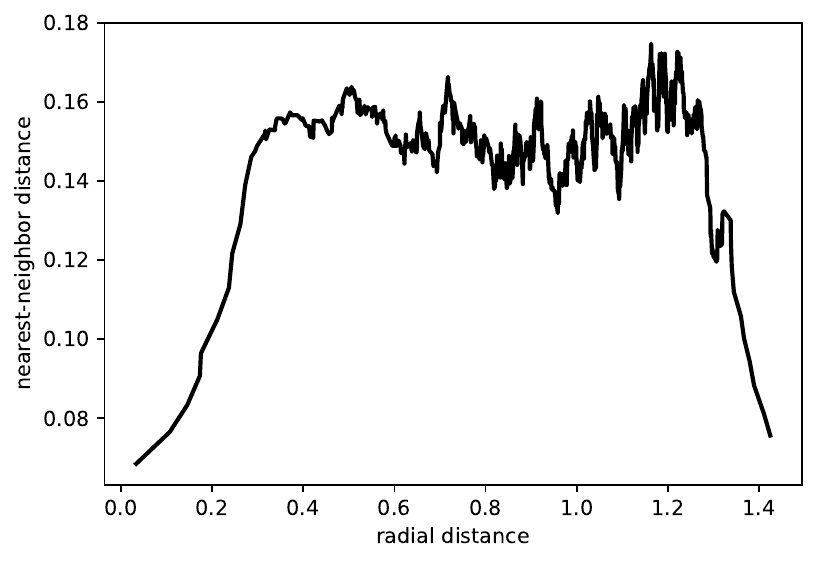}
	\caption{A plot of the nearest-neighbor distance as a function of the distance to the center of mass for the 1000-particle CC roughly $1.5\%$ above the absolute minimum of $V$. Note that the configuration is a static CC (equilibrium in shape space), therefore the plot represents a time-independent state, not a transient snapshot.}
	\label{fig4}
\end{figure}

\begin{figure}[ht!]
	\centering
	\includegraphics[width=\columnwidth]{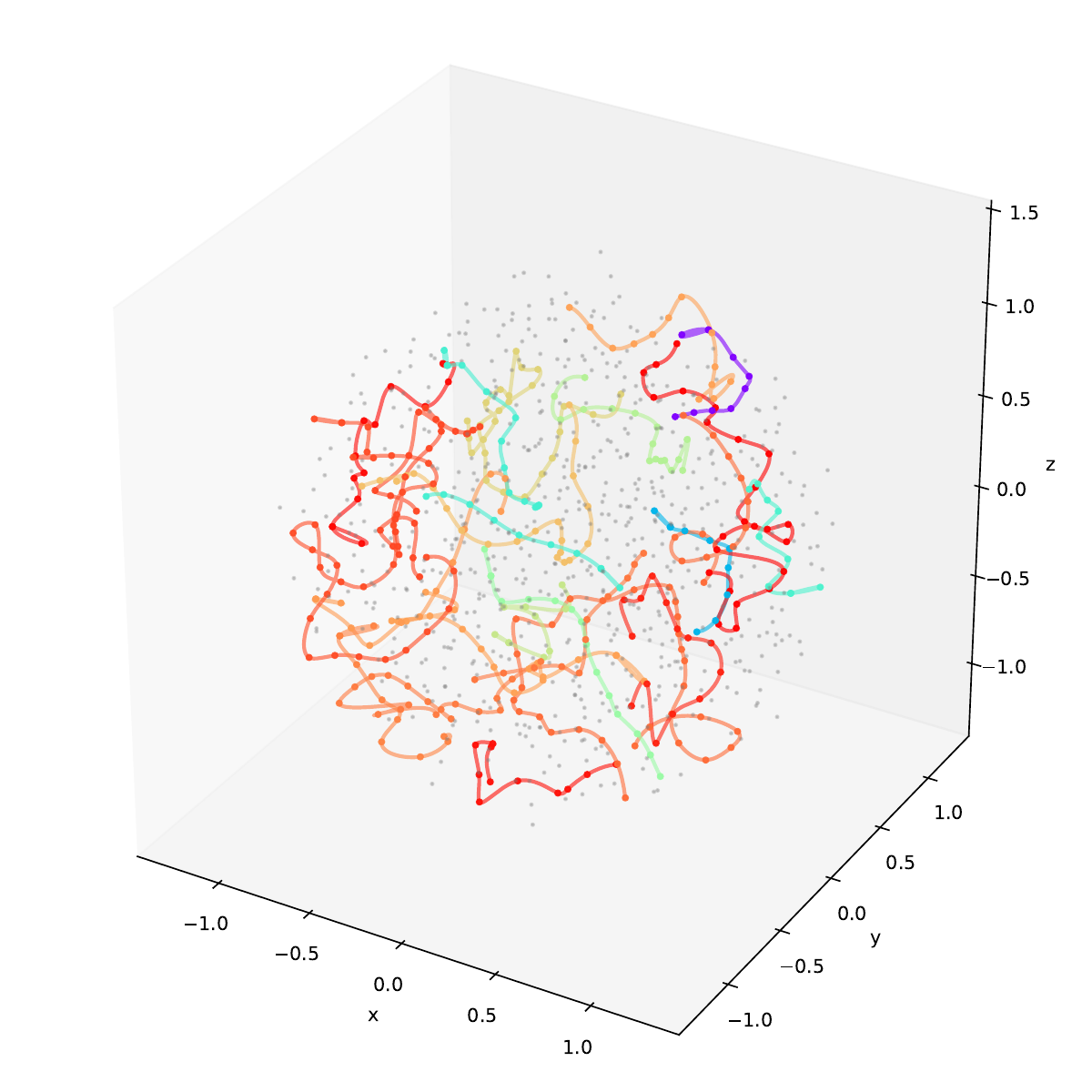}
	\caption{The previous 1000-particle CC roughly $1.5\%$ above the absolute minimum of $V$. A rainbow color scale is used to encode filament density, defined as the inverse of the mean nearest-neighbor distance, with purple indicating higher densities and red lower densities. Noise particles are shown in gray.}
	\label{fig5}
\end{figure}

Figures~\ref{fig2}--\ref{fig5} depict a three-dimensional 1000-particle CC with variety above the absolute minimum of $V$, given by Fig.~\ref{fig1}. In Fig.~\ref{fig3}, we color-code particles by their NN separation and in Fig.~\ref{fig4} we plot the NN spacing as a function of the radial distance from the center of mass. Finally, in Fig.~\ref{fig5} we visualize the formation of filaments in the CC above the absolute minimum of $V$. These simulations agree with the results presented in \cite{paper1}.
For equal-mass $N=1000$ central configurations, the absolute minimum of the variety is $V_{\text{min}} = 0.457916$. This value corresponds to the most uniform (least clustered) distribution we have found. The configuration analyzed here has variety $V = 0.465304$, i.e., roughly $1.5\%$ above the absolute minimum $V_{\text{min}}=0.457916$.

In the color-coded plot (Fig.~\ref{fig3}), where darker colors correspond to shorter distances to the nearest-neighbor, the predominance of green points indicates that most particles maintain an intermediate separation, while the presence of both blue and orange points suggests localized regions of closer and farther spacing, respectively. 

The plot of nearest-neighbor distance as a function of distance from the center of mass (Fig.~\ref{fig4}) shows that, for radial distances between $r\approx0.3$ and $r\approx1.3$, the NN distance fluctuates around a value of 0.16, with a slight increase toward the outer region, indicating a gradual loosening of the local particle packing. At smaller radii, particles are more densely packed, which means they experience stronger gravitational interactions, whereas particles at larger radii reside in shallower regions of the potential. The apparent decrease in NN separation for $r>1.3$ may be an artifact of low particle statistics. 

Furthermore, the filament analysis (Fig.~\ref{fig5}) indicates that a total of 24 elongated structures were detected, comprising 37.50\% of the particles. This suggests that the system exhibits partial anisotropic organization rather than being entirely homogeneous. 

Overall, these results point to a central configuration with a core--halo structure and the emergence of filamentary structures in a significant fraction of the system, consistent with hierarchical clustering patterns observed in gravitational systems. Such a core--halo structure is familiar in astrophysical systems such as globular clusters and dark-matter halos, but here it emerges in a purely static and scale-invariant Newtonian setting.

The radial variation of nearest-neighbor separations stands in clear tension with the assumption of large-scale homogeneity underlying the cosmological principle \cite{peebles1993principles}. Although this principle is intended as an approximation applicable at sufficiently large scales, central configurations show that even highly symmetric equilibrium solutions of Newtonian gravity generically exhibit spatially varying characteristic length scales. Homogeneity therefore does not arise automatically from gravitational dynamics but represents an idealization whose domain of validity must be explicitly assessed.

\section{Measured Geometry and the Philosophical Legacy of Poincaré and Einstein}\label{secIV}

The findings presented above gain full significance when interpreted through the philosophical and methodological perspectives of Henri Poincaré and Albert Einstein on the physical meaning of geometry. Both emphasized, in complementary ways, that the geometry of physical space cannot be understood independently of the material processes by which it is measured.

Poincaré was an early proponent of a fully operational view of geometry. In \textit{Science and Hypothesis} \cite{poincare1902}, he argued that the choice of the geometric framework---Euclidean or non-Euclidean---is not determined by pure reason but by empirical convenience, reflecting the behavior of measuring instruments. Because rods and clocks are physical systems subject to forces, their lengths and rates cannot be assumed invariant in all contexts. Geometry thus emerges not as an intrinsic attribute of space but as an effective, relational description inferred from the interactions between material bodies and measuring devices. Poincaré’s conventionalism replaces absolute geometry with a framework grounded in empirical and relational considerations.

Einstein extended and deepened this operational perspective in the formulation of general relativity. In his lecture \textit{Geometry and Experience} \cite{einstein1921}, he distinguished between axiomatic geometry, a purely mathematical construct, and practical geometry, defined by the physical behavior of rods and clocks. In the presence of gravitational fields, rods contract and clocks dilate, so that the geometry inferred from measurements depends directly on the distribution of mass--energy. Consequently, geometry becomes a dynamical entity, intrinsically linked to the physical laws governing matter and gravitation.

Although relativity endows spacetime with a continuous geometry described by a metric tensor, the underlying philosophical insight is independent of relativistic dynamics. Our numerical analysis of central configurations shows that the same operational reasoning holds within classical Newtonian gravity. The gravitational potential can be viewed as generating a classical analogue of geometric distortion, encoded not in a metric field but in the relational structure of particle separations. The systematic variation of nearest-neighbor distances indicates that idealized measuring rods, defined relationally through inter-particle separations, acquire lengths that depend on their position within the gravitational configuration. Therefore, dense regions correspond to shorter effective length standards, whereas dilute regions correspond to longer ones, implying a position-dependent geometry as inferred from measurement. In this sense, the geometry ``measured'' by these rods is inhomogeneous, despite the formally Euclidean nature of Newtonian space.

This distinction highlights an important conceptual separation between \emph{background geometry} and \emph{measured geometry}. Background geometry refers to the mathematical structure assumed in the formulation of the theory---in this case, Euclidean space. Measured geometry, by contrast, refers to the effective geometric relations inferred from physical measurements carried out by material systems embedded in that space. Our results show that measured geometry can depart significantly from the background geometry due solely to gravitational interactions, without invoking spacetime curvature in the relativistic sense.

The significance of this distinction is evident when noting that all empirical access to geometry is mediated by matter. Whether distances are measured with rods, clocks, light signals or particle separations, the process is constrained by the physical interactions acting on the measuring apparatus. In self-gravitating systems, gravity naturally produces spatial inhomogeneities in the matter distribution, which in turn shape the relational structure from which geometric properties are inferred. Consequently, the emergence of an effective, position-dependent geometry is not anomalous but a direct outcome of the operational definition of measurement. 

\section{Toward a Unified Framework for Emergent Measured Geometry in Gravitational Systems}\label{secV}

The results of this study indicate that measured geometry in gravitational systems is neither fixed nor fundamental, but emerges from the relational structure defined by matter distributions and their interactions. This viewpoint motivates the development of a unified conceptual framework that integrates discrete Newtonian systems, continuum gravitational theories and modern emergent-spacetime approaches within a single operationally grounded language.

Central to this framework is the recognition that all geometric information available to observers is mediated by physical processes. Distances, durations and angles are not abstract primitives, but inferred from the behavior of material systems---rods, clocks, light signals or, in the present context, inter-particle separations. Self-gravitating systems are necessarily shaped by gravity, rendering geometry intrinsically dependent on interactions. The inhomogeneous distributions revealed in central configurations therefore offer a minimal and concrete illustration of emergent measured geometry.

A natural connection between discrete and continuum descriptions is provided by geometric formulations of Newtonian gravity. In Newton--Cartan theory, Newtonian gravitation is expressed geometrically by replacing absolute space and time with a covariant spacetime structure defined via a connection compatible with a degenerate metric \cite{malament2012}. Although this framework is typically applied to smooth matter distributions, the discrete central configurations analyzed here indicate that an effective Newton--Cartan geometry may emerge through coarse-graining of relational particle data. Establishing systematic methods to derive continuum geometric structures from discrete gravitational equilibria constitutes a promising avenue for future research.

Connections also arise with contemporary approaches in which spacetime geometry and gravity are viewed as emergent phenomena. Entropic gravity models propose that gravitational dynamics originate from thermodynamic or informational principles, with geometry arising as an effective macroscopic description \cite{verlinde2011, verlinde2017}. Similarly, holographic frameworks suggest that spacetime geometry is encoded in non-geometric degrees of freedom. Although these approaches operate in regimes far removed from classical Newtonian gravity, the central configurations studied here exhibit a key shared feature: geometry is not fundamental but arises from underlying relational structure.

From a methodological perspective, discrete gravitational systems like central configurations provide a valuable testing ground for emergent-geometry concepts. Unlike full cosmological simulations or quantum gravity models, they are computationally tractable, conceptually transparent and grounded in well-understood physics. By systematically varying particle number, mass distribution and dimensionality, one can investigate how measured geometry depends on microscopic structure and interaction strength. Such studies may reveal robust, scale-independent features of emergent geometry that are likely to persist across diverse theoretical frameworks.

Finally, the perspective developed here invites a reorientation of foundational questions in gravitational physics. Instead of asking which geometry spacetime \emph{possesses}, one may ask how geometric relations \emph{emerge} from physical interactions and measurement processes. This shift extends the insights of Poincaré and Einstein into a contemporary, relational framework. By integrating discrete Newtonian models, geometric reformulations of gravity and emergent-spacetime paradigms under the concept of measured geometry, we advance toward a coherent understanding of space as a physical, dynamical and relational construct.

\section{Summary and Discussion}\label{secVI}

In this work we have investigated the emergence of measured geometry in self-gravitating systems from a relational, scale-invariant perspective. By studying central configurations of the Newtonian 
$N$-body problem, we showed that even equilibrium gravitational systems display systematic spatial inhomogeneities in inter-particle separations. Interpreted operationally as local measuring rods, these separations define an effective geometry that varies with position, despite the underlying formulation in absolute Euclidean space.

A central conceptual thread of our analysis is the primacy of scale invariance and the structure of shape space. Central configurations represent equilibria not in absolute space, but in shape space itself. Their defining feature---invariance under uniform scaling---naturally elevates dimensionless, relational quantities to the status of physically meaningful observables. Scale invariance implies that no external length scale governs the structure of the system; instead, characteristic scales arise dynamically from the matter distribution. From this perspective, geometry is most appropriately understood in terms of scale-free relations among particles, rather than absolute lengths defined against a background.

This viewpoint naturally extends to cosmology, where observations rely almost entirely on dimensionless, relational quantities such as redshifts, angular sizes and clustering statistics. The inhomogeneous geometries associated with central configurations suggest that aspects of cosmological phenomenology may admit a relational, scale-invariant reinterpretation.

Future research could explore whether dimensionless measures of structure---such as clustering statistics or scale-invariant complexity measures like the variety---offer alternative ways to characterize cosmic evolution. It may also be worth investigating how large-scale inhomogeneities influence the operational inference of cosmological parameters, potentially shedding light on effects often attributed to dark energy or cosmic acceleration. Although speculative, these results indicate that gravitational systems can produce effective geometric variation without introducing new physical ingredients.

Note that although the configurations studied here are static equilibria, they serve as initial conditions for time-dependent $N$-body simulations. The characteristic collapse time for a uniform, pressureless sphere is the free-fall time $t_{\text{ff}} = \sqrt{3\pi/(32G\bar\rho)}$ (with $\bar\rho$ the mean density) [see, e.g., \cite{BinneyTremaine2008}, §2.2]; a small perturbation of our equilibrium configurations would lead to gravitational collapse on this timescale. We plan to investigate such dynamical evolutions in a future study.

Equally significant are the implications for quantum gravity. Many contemporary approaches, including loop quantum gravity, causal set theory and holographic frameworks, posit that spacetime geometry is not fundamental but emerges from more primitive, non-geometric degrees of freedom. Shape space provides a natural classical precursor to this idea: it is a configuration space defined entirely by relational data, free of absolute scale. The emergence of effective geometry from shape dynamics in Newtonian gravity suggests that relational and scale-invariant structures may play a foundational role even in the quantum regime.

In particular, the absence of an absolute length scale in shape space aligns with approaches to quantum gravity in which scale is not fundamental but emergent. Developing quantization schemes directly on shape space or identifying quantum counterparts of the variety offers a promising avenue for research. In this context, discrete gravitational systems like the $N$-body problem may serve as useful toy models for exploring the emergence of geometry in quantum gravity.

More broadly, these results support a shift from viewing geometry as a primitive arena to treating it as a derived entity arising from relational structure and interaction. In this sense, the framework extends the philosophical insights of Poincaré and Einstein into a concrete relational setting. Central configurations show that this viewpoint yields substantive consequences even within classical physics.

In conclusion, scale invariance and shape space provide a unified framework for understanding the emergence of measured geometry in gravitational systems. By demonstrating how effective geometry arises from relational structure in a simple classical model, this work lays conceptual foundations for further developments in Newtonian dynamics, cosmology and quantum gravity.

\section*{Acknowledgments}
MIRL acknowledges the Instituto de Astrofísica e Ciências do Espaço (IA) for providing the computer cluster used in this work. JB thanks Tim Koslowski for many helpful discussions. FSNL acknowledges support from the Fundação para a Ciência e a Tecnologia (FCT) through a Scientific Employment Stimulus contract (reference CEECINST/00032/2018), as well as funding from the research grants UID/04434/2025 and PTDC/FIS-AST/0054/20.

\bibliographystyle{plain}

\begin{thebibliography}{99}
	
	\bibitem{Newton1687}
	I. Newton,
	\textit{Philosophiæ Naturalis Principia Mathematica},
	London, 1687.

    \bibitem{malament2012}
	D. Malament,
	\textit{Topics in the Foundations of General Relativity and Newtonian Gravitation Theory},
	University of Chicago Press, 2012.

    \bibitem{verlinde2011}
	E. Verlinde,
	\textit{On the Origin of Gravity and the Laws of Newton},
	JHEP \textbf{1104}, 029 (2011), \href{https://arxiv.org/abs/1001.0785}{arXiv:1001.0785}.
	
	\bibitem{verlinde2017}
	E. Verlinde,
	\textit{Emergent Gravity and the Dark Universe},
	SciPost Phys. \textbf{2}, 016 (2017), \href{https://arxiv.org/abs/1611.02269}{arXiv:1611.02269}.

    \bibitem{paper1}
    M. I. R. Lourenço, J. Barbour, F. S. N. Lobo, 
    \textit{Scale Invariance, Variety and Central Configurations} (2026), \href{https://arxiv.org/abs/2602.11225}{arXiv:2602.11225}.

    \bibitem{moulton1910}
	F. R. Moulton,
	\textit{The straight line solutions of the problem of $n$ bodies},
	Annals of Mathematics \textbf{12}(1), 1–17 (1910).
	
	\bibitem{saari_cc} 
	  D. G. Saari, 
    \textit{On the role and the properties of $N$-body central configurations},
    Celestial Mechanics \textbf{21}, 9–20 (1980). 
	
	\bibitem{meyer2009}
	K. Meyer, G. Hall, D. Offin,
	\textit{Introduction to Hamiltonian Dynamical Systems and the $N$-Body Problem},
	2nd edition, Springer, 2009.

    \bibitem{peebles1993principles}
	P. J. E. Peebles,
	\textit{Principles of Physical Cosmology},
	Princeton University Press, 1993.

	%\bibitem{einstein1916foundation}
	%A. Einstein,
	%\textit{The Foundation of the General Theory of Relativity},
	%Annalen der Physik \textbf{49}, 769–822 (1916).
    
	\bibitem{poincare1902}
	H. Poincaré,
	\textit{Science and Hypothesis},
	Dover Publications, 1952 (original 1902).
	
	\bibitem{einstein1921}
	A. Einstein,
	\textit{Geometry and Experience},
	lecture delivered in 1921, reprinted in:
	A. Einstein,
    \textit{Sidelights on Relativity},
	Princeton University Press, 1922.
    	
    \bibitem{BinneyTremaine2008}
    J.~Binney and S.~Tremaine, \textit{Galactic Dynamics}, 2nd ed.
    (Princeton University Press, Princeton, N.J., 2008).	
    	
	%\bibitem{weinberg2008cosmology}
	%S. Weinberg,
	%\textit{Cosmology},
	%Oxford University Press, 2008.
	
	%\bibitem{bond1996filaments}
	%J. R. Bond, L. Kofman, D. Pogosyan,
	%\textit{How filaments of galaxies are woven into the cosmic web},
	%Nature, \textbf{380}, 603–606 (1996).
	
	%\bibitem{springel2005simulations}
	%V. Springel et al.,
	%\textit{Simulations of the formation, evolution and clustering of galaxies and quasars},
	%Nature, \textbf{435}, 629–636 (2005).
	
\end{thebibliography}

\end{document}